\documentclass[pra,twocolumn,amsmath,amssymb,superscriptaddress]{revtex4-1}
\usepackage{epsfig,amsmath}
\usepackage{subfigure}
\usepackage{graphicx}% Include figure files
\usepackage{dcolumn}% Align table columns on decimal point
\usepackage{stmaryrd}
\usepackage{mathrsfs}
\usepackage{pifont}
\usepackage{amsthm}
\usepackage{amssymb}
\usepackage{bm}
\usepackage{latexsym}
\usepackage[colorlinks=true,linkcolor=blue,citecolor=blue]{hyperref}
\usepackage{color}
\usepackage{epstopdf}

\begin{document}

\title{Generating entangled states from coherent states in circuit-QED}

\author{Shi-fan Qi}
\affiliation{School of Physics, Zhejiang University, Hangzhou 310027, Zhejiang, China}

\author{Jun Jing}
\email{jingjun@zju.edu.cn}
\affiliation{School of Physics, Zhejiang University, Hangzhou 310027, Zhejiang, China}

\date{\today}

\begin{abstract}
Entangled states are self-evidently important to a wide range of applications in quantum communication and quantum information processing. We propose an efficient and convenient two-step protocol for generating Bell states and NOON states of two microwave resonators from merely coherent states. In particular, we derive an effective Hamiltonian for resonators coupled to a superconducting $\Lambda$-type qutrit in the dispersive regime. By the excitation-number-dependent Stark shifts of the qutrit transition frequencies, we are able to individually control the amplitudes of specified Fock states of the resonators associated with relevant qutrit transition, using carefully tailored microwave drive signals. Thereby an arbitrary bipartite entangled state in Fock space can be generated by a typical evolution-and-measurement procedure. We analysis the undesired state transitions and the robustness of our protocol against the systematic errors from the microwave driving intensity and frequency, the quantum decoherence of all components, and the crosstalk of two resonators. In addition, we demonstrate that our protocol can be extended to a similar scenario with a $\Xi$-type qutrit.
\end{abstract}

\date{\today}

\maketitle

\section{Introduction}

Entangled states~\cite{qe} of harmonic oscillators, e.g., microwave resonators, play a key role in quantum technologies, such as quantum communication~\cite{qco} and quantum information processing~\cite{qip}. Creation, manipulation, and measuring of the entangled states in both experimental platforms~\cite{trapion,ghzstate,superconduct2,superconduct3,Schrodingerstate} and theoretical protocols have therefore been intensively pursued for a long time~\cite{GHZ,ghzstate,GHZstate2,bellstate,bellstate3,noon1,noon2,wernerstate} and are still under an active investigation.

The simplest and yet the most popular maximally entangled states in the Fock space of the resonators are Bell states, i.e., $(|00\rangle\pm|11\rangle)/\sqrt{2}$ and $(|10\rangle\pm|01\rangle)/\sqrt{2}$, where $|0\rangle$ and $|1\rangle$ are the ground and the first excited states, respectively. Generating Bell states in photonic systems is fundamental to both quantum cryptography~\cite{cryptography} and quantum teleportation~\cite{teleportation}. Another widely applied entangled states are the NOON states $(|N0\rangle\pm|0N\rangle)/\sqrt{2}$ with $N$ integer, consisting of two orthogonal components in an equal-weighted superposition~\cite{noonstate,noon8,noon9,noon10,floquetnoon}. They are crucial elements in quantum metrology~\cite{quanmea,quantmetro}, quantum optical lithography~\cite{quaninter,quanlith}, and quantum information processing~\cite{qip}. The NOON states have been realized in multiple quantum systems, including polarization states of photons~\cite{polarnoon}, nuclear spin of molecules~\cite{magnetsense}, optical paths of photons~\cite{lightnoon}, ultracold dipolar atoms in optical superlattice setup~\cite{atominter,designnoon}, and phonons in ion trap~\cite{ionnoon}. Nevertheless, the multi-step ultra-precise control over the quantum devices, the specialization of the initial states, and the decoherence of quantum systems make it hard to create and hold highly-entangled states. Fast and convenient protocols for preparing either Bell state or NOON state are still underway.

With unique properties including long coherent time~\cite{fq,fq2,fq3} and strong and even ultrastrong dipole-dipole coupling~\cite{sq,sq2,ulstrong,ulstrong2,cq}, the circuit-QED system~\cite{sq3,circuitqed} has been used as a promising platform to generate nonclassical states and entangled states~\cite{nonclass,ghzstate,sq,sq2}. Many protocols have thus been proposed upon the manipulation capability up to the artificial atomic level. It is noted that most of the existing generation protocols~\cite{bellstate,bellstate3,noon1,noon2,noon3,noon4,noon5,noon6,noon7} are developed on Rabi oscillations and initialization of the resonator. For example, by tuning the atomic frequency to be resonant with the resonator, the excitation of the artificial atom is transferred to the resonator mode through a half of Rabi oscillation. The NOON state with a significant number $N$ is then generated by a step-by-step processing with resonators prepared as the ground state, that is fragile to both systematic error and environmental noise.

\begin{figure}[htbp]
\centering
\includegraphics[width=0.45\textwidth]{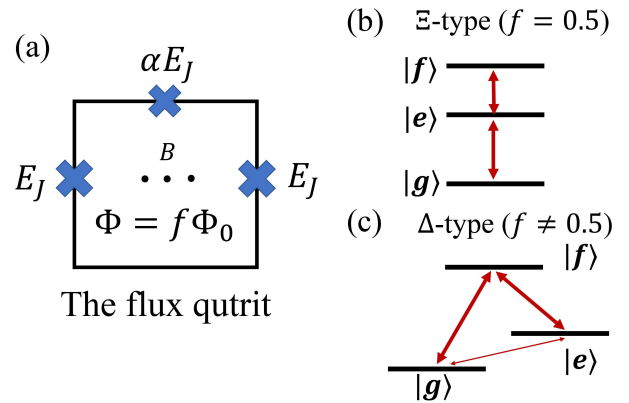}
\caption{(a): A superconducting loop interrupted by three Josephson junctions. Two of them have the same coupling energy $E_J$ and the third has a smaller one with $0.5<\alpha<1$. The loop is biased by a magnetic field $B$. $\Phi=f\Phi_0$ is the magnetic flux penetrating the loop, where $\Phi_0$ is the magnetic-flux quantum. (b) and (c): Lowest levels' configuration for the $\Xi$-type qutrit with $f=0.5$ and that for the $\Delta$-type qutrit with $f\neq0.5$, respectively. }\label{diagram1}
\end{figure}

In this work, we propose a two-step protocol for entangling two resonators that are dispersively coupled to a superconducting qutrit~\cite{sq,sq2,sq3}. As shown in Fig.~\ref{diagram1}, the flux-driven qutrit~\cite{sq3} presents various configuration by the magnitude of the magnetic flux. At a particular point with $f=0.5$, it behaves as a $\Xi$-type qutrit. When $f\neq0.5$, the parity symmetry is broken; all three dipole transitions among $|g\rangle$, $|e\rangle$, and $|f\rangle$ are allowed; and the qutrit becomes a $\Delta$-type one. The resonators are initially in separable coherent states, which can be conveniently created by classical drives~\cite{concoh}. The main ingredients of our protocol are the excitation-number-dependent qutrit rotations, allowing individually manipulating the probability amplitudes of the Fock-state components of coherent states by carefully tailoring the microwave drive signals. The excitation-number-dependent Stark shift in the qutrit transition frequency arises from the qutrit-resonator interaction~\cite{dispersive1,dispersive2}. It has been used to realize dressed coherent states~\cite{dress}, entangled states in discrete or discrete-continuous systems~\cite{atomentan,entangle2}, quantum phase gate~\cite{phasegate,phasegate2}, and hybrid Fredkin gate~\cite{Fredkin}. Experimentally it was reported to generate arbitrary Fock states~\cite{concoh} in circuit-QED system.

In Sec.~\ref{model}, we introduce our model of two resonators coupled to a $\Lambda$-type qutrit and derive the effective Hamiltonian at the dispersive regime. In Sec.~\ref{prepar}, we provide a detailed protocol for generating a bipartite entangled state in Fock space. Appendix~\ref{appa}, we discuss the nonideal measurement. We analysis in Sec.~\ref{detunfide} the final state fidelity in the presence of unsuppressed state-transitions. The detailed calculation is provided in Appendix~\ref{appb}. The systematic errors from the driving intensity and frequency, the decoherence noises for the overall qutrit-resonator system, and the unwanted couplings in the whole system are respectively addressed in Sec.~\ref{sysfide}, Sec.~\ref{decofide}, and Sec.~\ref{uncoupling}. In Sec.~\ref{discuss}, our protocol is extended to the $\Xi$-type transmon qutrit. The whole work is summarized in Sec.~\ref{conclu}.

\section{Model and the effective Hamiltonian}\label{model}

Consider a $\Delta$-type qutrit as shown in Fig.~\ref{diagram1}(c) coupled to two microwave resonators~\cite{circuitqed} (labeled $a$ and $b$). The three levels of the qutrit $|g\rangle$, $|e\rangle$, and $|f\rangle$ denote respectively the ground state, the intermediate state, and the highest-level state. The system Hamiltonian ($\hbar=1$) is given by
\begin{equation}\label{deltaHamiltonian}
\begin{aligned}
H&=H_0+V,\\
H_0&=\omega_a a^{\dag}a+\omega_b b^{\dag}b+\omega_e|e\rangle\langle e|+\omega_{f}|f\rangle\langle f|,\\
V&=[g_{eg}^a(a^{\dag}+a)+g_{eg}^b(b+b^{\dag})](|e\rangle\langle g|+|g\rangle\langle e|)\\
&+[g_{fg}^a(a^{\dag}+a)+g_{fg}^b(b+b^{\dag})](|f\rangle\langle g|+|g\rangle\langle f|)\\
&+[g_{fe}^a(a^{\dag}+a)+g_{fe}^b(b+b^{\dag})](|f\rangle\langle e|+|e\rangle\langle f|).
\end{aligned}
\end{equation}
where $\omega_a$ ($\omega_b$) is the transition frequency of the resonator $a$ ($b$), $\omega_{e}$ ($\omega_{f}$) represents the free level spacing between the state $|e\rangle$ ($|f\rangle$) and the ground state.

With proper parameters in junctions, the elements in the transition matrix describing $|f\rangle\leftrightarrow|e\rangle$ and $|f\rangle\leftrightarrow|g\rangle$ are dominant over that for $|e\rangle\leftrightarrow|g\rangle$ in magnitude, i.e., $g_{fg}, g_{fe}\gg g_{eg}$~\cite{sq3}. The $\Delta$-type qutrit could thus be simplified as the $\Lambda$-type. Through frequency manipulation, the resonator $a$ ($b$) can only be significantly coupled to the $|e\rangle\leftrightarrow|f\rangle$ ($|g\rangle\leftrightarrow|f\rangle$) transition. The two transitions are assumed to be sufficiently detuned or decoupled from each other and then these two couplings contribute to independent Rabi transitions. Thus, the interaction Hamiltonian $V$ in Eq.~\eqref{deltaHamiltonian} can be approximated as
\begin{equation}\label{Hamiltonian}
\begin{aligned}
V&=g_a(|f\rangle\langle e|+|e\rangle\langle f|)(a+a^{\dag})\\
&+g_b(|f\rangle\langle g|+|g\rangle\langle f|)(b+b^\dag),
\end{aligned}
\end{equation}
where $g_a\equiv g_{fe}^a$ and $g_b\equiv g_{fg}^b$.

In the dispersive regime, i.e., $g_a\ll|\omega_f-\omega_e-\omega_a|$ and $g_b\ll|\omega_f-\omega_b|$, $V$ can be considered as a perturbation with respect to the free Hamiltonian $H_0$. Across the whole Hilbert space, the interaction Hamiltonian would give rise to the energy shift of any eigenstate $|i\rangle$ of the unperturbed Hamiltonian $H_0$ in Eq.~(\ref{deltaHamiltonian}). To the second order, the energy shift is effectively given by
\begin{equation}\label{chi}
\chi=\sum_{j\ne i}\frac{V_{ij}V_{ji}}{\omega_i-\omega_j},
\end{equation}
where $V_{ji}\equiv\langle j|V|i\rangle$ and $\omega_{i}$ is the eigenenergy of $|i\rangle$. For $H_0$, the eigenstates read $|i\rangle=|gnm\rangle\equiv|g\rangle|n\rangle_a|m\rangle_b$, $|enm\rangle$, and $|fnm\rangle$.

Summarizing the two paths starting from $|gnm\rangle$ and going back to $|gnm\rangle$ through an intermediate state, i.e., $|gnm\rangle\to|fn(m+1)\rangle\to|gnm\rangle$ and $|gnm\rangle\to|fn(m-1)\rangle\to|gnm\rangle$, one can obtain the second-order correction in energy for the state $|gnm\rangle$,
\begin{equation}\label{energyg}
n\omega_a+m\omega_b-m\chi_b-(m+1)\chi'_b
\end{equation}
according to Eq.~(\ref{chi}). In the same way, one can obtain the corrected energy for the state $|enm\rangle$
\begin{equation}\label{energye}
\omega_e+n\omega_a+m\omega_b-n\chi_a-(n+1)\chi'_a,
\end{equation}
and the corrected energy for the state $|fnm\rangle$
\begin{equation}\label{energyf}
\omega_f+n\omega_a+m\omega_b+(n+1)\chi_a+n\chi'_a+(m+1)\chi_b+m\chi'_b,
\end{equation}
where
\begin{equation}\label{chiab}
\begin{aligned}
\chi_a&=\frac{g^2_a}{\omega_f-\omega_e-\omega_a},\quad \chi'_a=\frac{g^2_a}{\omega_f-\omega_e+\omega_a},\\
\chi_b&=\frac{g^2_b}{\omega_f-\omega_b},\quad \chi'_b=\frac{g^2_b}{\omega_f+\omega_b},
\end{aligned}
\end{equation}
are the energy shifts due to the interaction Hamiltonian $V$, describing the virtual two-photon processes in Eq.~(\ref{chi}). Note $\chi$'s and $\chi'$'s are contributions from rotating-wave interaction and counterrotating terms, respectively. The presence or absence of the counterrotating terms does not violate the following derivation and protocol.

It is noted that $|n\rangle$ and $|m\rangle$ are arbitrarily Fock states in the preceding analysis. Thus, in the dispersive regime, the effective Hamiltonian can be written as
\begin{equation}\label{Heff}
\begin{aligned}
H_{\rm eff}&=-\chi_b'|g\rangle\langle g|+\left(\omega_e-\chi'_a\right)|e\rangle\langle e|\\
&+\left(\omega_f+\chi_a+\chi_b\right)|f\rangle\langle f|+\omega_a a^{\dag}a+\omega_b b^\dag b\\
&+\left(\chi_a+\chi'_a\right)a^{\dag}a(|f\rangle\langle f|-|e\rangle\langle e|)\\
&+\left(\chi_b+\chi'_b\right)b^{\dag} b(|f\rangle\langle f|-|g\rangle\langle g|),
\end{aligned}
\end{equation}
where the last two lines describe the excitation-number-dependent Stark shifts for all the three atomic levels.

\section{Two-step generation protocol}\label{prepar}

\begin{figure}[htbp]
\centering
\includegraphics[width=0.45\textwidth]{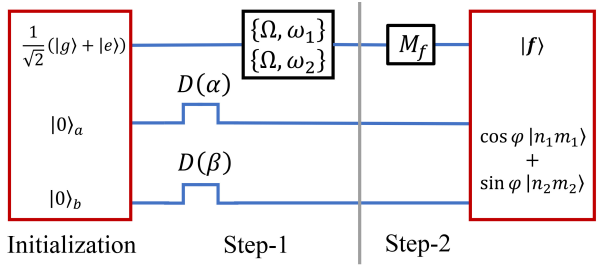}
\caption{Pulse sequence and measurement for the entangled state generation. The qutrit is initially in a balanced superposed state. $D(\alpha)$ and $D(\beta)$ are displacement operation to prepare the resonators into coherent states. In Step-1, two microwave pulses $\{\Omega, \omega_1\}$ and $\{\Omega, \omega_2\}$ are simultaneously applied to qutrit. In Step-2, a projective measurement $M_f$ is imposed on $|f\rangle$ to yield the final entangled state.}\label{diagram2}
\end{figure}

In this section, we show how to generate a bipartite entangled state of two resonators in the form of $(|n_1m_1\rangle+|n_2m_2\rangle)/\sqrt{2}$, $n_1\neq n_2$ and $m_1\neq m_2$, from coherent states by the effective Hamiltonian in Eq.~(\ref{Heff}). The whole protocol is visually summarized in Fig.~\ref{diagram2}.

Suppose that the qutrit is initially in the superposed state $(|g\rangle+|e\rangle)/\sqrt{2}$ and the two resonators $a$ and $b$ are prepared in the coherent states $|\alpha\rangle$ and $|\beta\rangle$, respectively. The whole system starts from
\begin{equation}\label{varstate0}
\begin{aligned}
|\Phi(0)\rangle&=\frac{1}{\sqrt{2}}(|g\rangle+|e\rangle)\otimes|\alpha\rangle\otimes|\beta\rangle, \\
|\alpha\rangle&=e^{-\frac{1}{2}|\alpha|^2}\sum^{\infty}_{n=0}\frac{\alpha^n}{\sqrt{n!}}|n\rangle, \\
|\beta\rangle&=e^{-\frac{1}{2}|\beta|^2}\sum^{\infty}_{m=0}\frac{\beta^m}{\sqrt{m!}}|m\rangle,
\end{aligned}
\end{equation}
where $\alpha$ and $\beta$ are complex numbers. Alternatively, the initial state can be written as
\begin{equation}\label{initialstate}
|\Phi(0)\rangle=\mathcal{N}\sum^{\infty}_{n,m}\frac{\alpha^n\beta^m}{\sqrt{n!m!}}(|gnm\rangle+|enm\rangle),
\end{equation}
where $\mathcal{N}=\exp[-(|\alpha|^2+|\beta|^2)/2]/\sqrt{2}$ is the normalization coefficient. The initial coherent state can be achieved by the application of a microwave pulse on the resonator, that produces a displacement operation $D(\alpha)\equiv\exp(\alpha a^\dag-\alpha^*a)$ on the vacuum state $|0\rangle$ and $|\alpha\rangle=D(\alpha)|0\rangle$~\cite{concoh}.

Step-$1$: Applying two microwave pulses $\{\Omega,\omega_1\}$ and $\{\Omega,\omega_2\}$ that are nearly resonant with the qutrit transitions $|e\rangle\leftrightarrow |f\rangle$ and $|g\rangle\leftrightarrow |f\rangle$, respectively. Here $\Omega$ is the intensity of the driving pulses satisfying $\Omega T=\pi/2$ with the duration time $T$ and $\omega_j$, $j=1,2$, is the driving frequency. The total Hamiltonian including the driving pulses can then be written as
\begin{equation}\label{Htotorigin}
\begin{aligned}
H_{\rm tot}&=H_{\rm eff}+H_d,\\
H_d&=\Omega\left(|f\rangle\langle e|e^{-i\omega_1t}+|f\rangle\langle g|e^{-i\omega_2t}+\rm{H.c.}\right).
\end{aligned}
\end{equation}
In the rotating frame with respect to $U(t)=\exp(iH_{\rm eff}t)$, it turns out to be
\begin{equation}\label{Htotinter}
\begin{aligned}
H'_{\rm tot}&=U(t)H_{\rm tot}U^\dag(t)-iU(t)\dot{U}^\dag(t) \\
&=\Omega\left[|f\rangle\langle e|e^{i(\omega_{fe}-\omega_1)t}+|f\rangle\langle g|e^{i(\omega_{fg}-\omega_2)t}+{\rm H.c.}\right],\\
\omega_{fe}&=\omega_f-\omega_e+(2a^\dag a+1)(\chi_a+\chi'_a)\\ &+b^\dag b(\chi_b+\chi'_b)+\chi_b,\\
\omega_{fg}&=\omega_f+(2b^\dag b+1)(\chi_b+\chi'_b)+a^\dag a(\chi_a+\chi'_a)+\chi_a.
\end{aligned}
\end{equation}
Consequently, if one chooses
\begin{equation}\label{omega}
\begin{aligned}
\omega_1&=\omega_f-\omega_e+(2n_1+1)(\chi_a+\chi'_a)+m_1(\chi_b+\chi'_b)+\chi_b,\\
\omega_2&=\omega_f+(2m_2+1)(\chi_b+\chi'_b)+n_2(\chi_a+\chi'_a)+\chi_a,
\end{aligned}
\end{equation}
then in the Fock space, the total Hamiltonian could be rewritten as
\begin{equation}\label{Htot}
\begin{aligned}
H'_{\rm tot}&=\Omega\left(|fn_1m_1\rangle\langle en_1m_1|+|fn_2m_2\rangle\langle gn_2m_2|\right)\\
&+\Omega\sum_{n\ne n_1,m\ne m_1}^{\infty}|fnm\rangle\langle enm|e^{i\Delta_{nm}t}\\
&+\Omega\sum_{n\ne n_2,m\ne m_2}^{\infty}|fnm\rangle\langle gnm|e^{i\Delta'_{nm}t}+\rm{H.c.},
\end{aligned}
\end{equation}
where the nonvanishing detunings read
\begin{equation}\label{delta12}
\begin{aligned}
\Delta_{nm}&=\omega_f-\omega_e+(2n+1)(\chi_a+\chi'_a)\\
&+m(\chi_b+\chi'_b)+\chi_b-\omega_1,\\
\Delta'_{nm}&=\omega_f+(2m+1)(\chi_b+\chi'_b)+n(\chi_a+\chi'_a)+\chi_a-\omega_2.
\end{aligned}
\end{equation}

In the ideal weak-driving regime, these detunings $\Delta_{nm}$ and $\Delta'_{nm}$ are much larger than the Rabi frequency, i.e., $\Delta_{nm}, \Delta'_{nm}\gg\Omega$. In this case, the first driving pulse $\{\Omega, \omega_1\}$ resonantly drives the qutrit transition $|e\rangle\leftrightarrow|f\rangle$ and the number state $|n_1m_1\rangle$ of the resonator modes and its impact on the other number states could be averaged out in a properly long time-scale $\gg1/\Delta_{nm}$. Similarly, the second driving pulse $\{\Omega, \omega_2\}$ resonantly drives the state transition $|gn_2m_2\rangle\leftrightarrow|fn_2m_2\rangle$ and also does not significantly affect the number states other than $|n_2m_2\rangle$. Driven by the total Hamiltonian in Eq.~(\ref{Htot}) with secular approximation, the initial state of the system $|\Phi(0)\rangle$ evolves as
\begin{equation}\label{phit}
\begin{aligned}
|\Phi(t)\rangle&\approx\mathcal{N}\frac{\alpha^{n_1}\beta^{m_1}}{\sqrt{n_1!m_1!}}
\left[\cos(\Omega t)|e\rangle-i\sin(\Omega t)|f\rangle\right]|n_1m_1\rangle\\
&+\mathcal{N}\frac{\alpha^{n_2}\beta^{m_2}}{\sqrt{n_2!m_2!}}
\left[\cos(\Omega t)|g\rangle-i\sin(\Omega t)|f\rangle\right]|n_2m_2\rangle\\
&+\mathcal{N}\sum^{\infty}_{n\neq n_1,m\neq m_1}\frac{\alpha^n\beta^m}{\sqrt{n!m!}}|enm\rangle\\
&+\mathcal{N}\sum^{\infty}_{n\neq n_2,m\neq m_2}\frac{\alpha^n\beta^m}{\sqrt{n!m!}}|gnm\rangle.
\end{aligned}
\end{equation}
After a period of $\Omega T=\pi/2$, we have
\begin{equation}\label{phiTt}
\begin{aligned}
|\Phi(T)\rangle&=-i\mathcal{N}\left(\frac{\alpha^{n_1}\beta^{m_1}}{\sqrt{n_1!m_1!}}|fn_1m_1\rangle
+\frac{\alpha^{n_2}\beta^{m_2}}{\sqrt{n_2!m_2!}}|fn_2m_2\rangle\right)\\
&+\mathcal{N}\sum^{\infty}_{n\neq n_1,m\neq m_1}\frac{\alpha^n\beta^m}{\sqrt{n!m!}}|enm\rangle\\
&+\mathcal{N}\sum^{\infty}_{n\neq n_2,m\neq m_2}\frac{\alpha^n\beta^m}{\sqrt{n!m!}}|gnm\rangle.
\end{aligned}
\end{equation}

The two tailored microwave pulses could be separably applied to the system. One can first drive the transition $|e\rangle\leftrightarrow|f\rangle$ by the microwave pulse $\{\Omega, \omega_1\}$. The total Hamiltonian under the secular approximation is thus described by $\Omega|fn_1m_1\rangle\langle en_1m_1|+$H.c.. Then after a period $\Omega T=\pi/2$, the state $|en_1m_1\rangle$ evolves into $|fn_1m_1\rangle$ without invoking the other transitions. Next another microwave pulse $\{\Omega, \omega_2\}$ is applied to transfer $|gn_2m_2\rangle$ to $|fn_2m_2\rangle$. The state in Eq.~(\ref{phiTt}) is eventually achieved.

Step-$2$: Performing the measurement on the qutrit by the projection operator $M_f\equiv|f\rangle\langle f|$~\cite{measure1,measure2}. Consequently, the reduced density operator of two resonator modes $\rho_s$ becomes
\begin{equation}\label{rhos}
\rho_s(T)={\rm Tr}_q\left[\frac{M_f|\Phi(T)\rangle\langle\Phi(T)|M_f}{{\rm Tr}[M_f|\Phi(T)\rangle\langle\Phi(T)|M_f]}\right]=|\phi\rangle\langle\phi|,
\end{equation}
where the denominator $P={\rm Tr}[M_f|\Phi(T)\rangle\langle\Phi(T)|M_f]$ is the success probability~\cite{measure4,measure3} and ${\rm Tr}_q$ means partial trace over the qutrit. Up to a dynamical phase accumulated in evolution, $|\phi\rangle$ is an entangled state fully determined by $n_{1,2}$ and $m_{1,2}$ specified in Eq.~(\ref{omega}),
\begin{equation}\label{phi}
|\phi\rangle=\cos\varphi|n_1m_1\rangle+\sin\varphi|n_2m_2\rangle,
\end{equation}
where
\begin{equation}\label{theta}
\tan\varphi=\frac{\alpha^{n_2}\beta^{m_2}\sqrt{n_1!m_1!}}{\alpha^{n_1}\beta^{m_1}\sqrt{n_2!m_2!}}.
\end{equation}
A balanced superposition requires that $\tan\varphi=1$. Hereafter both $\alpha$ and $\beta$ are supposed to be positive numbers for simplicity. Up to a local phase, we have generated an arbitrary entangled state for two continuous-variable modes with a success probability
\begin{equation}\label{succproba}
P=\mathcal{N}^2\left(\frac{\alpha^{2n_1}\beta^{2m_1}}{n_1!m_1!}+\frac{\alpha^{2n_2}\beta^{2m_2}}{n_2!m_2!}\right)
\end{equation}
according to Eqs.~(\ref{phiTt}) and (\ref{rhos}).

Here are two distinguishing instances. With $n_1=m_1=0$ and $n_2=m_2=1$, one can obtain a double-excitation Bell state
\begin{equation}\label{bellstate}
|\phi\rangle=\frac{1}{\sqrt{2}}(|00\rangle+|11\rangle)
\end{equation}
by setting the coherent-state parameters as $\alpha=\beta=1$. With $n_1=m_2=0$, $m_1=n_2=N$, and $\alpha=\beta$, the entangled state in Eq.~(\ref{phi}) turns out to be a NOON state
\begin{equation}\label{Noonstate}
|\phi\rangle=\frac{1}{\sqrt{2}}(|0N\rangle+|N0\rangle).
\end{equation}

The efficiency of our protocol for generating $|\phi\rangle$ can be quantified by the state fidelity
\begin{equation}\label{fidelity}
F=\frac{\langle\phi|\langle f|\rho(T)|f\rangle|\phi\rangle}{{\rm Tr}[M_f\rho(T)M_f]},
\end{equation}
where $\rho(T)$ is the density matrix directly obtained by Eq.~(\ref{Htot}) without secular approximation. We also ignore the dynamical phase of the time-evolved state in practical numerical simulation. The protocol depends on a perfect performance by measurement operator $M_f$. The influence from a nonideal measurement on the state fidelity is shown in Appendix~\ref{appa}. We stick to the ideal measurement $M_{f}$ in the following analysis about fidelity.

\section{Fidelity analysis}\label{fide}

\subsection{Undesired transition in the presence of driving}\label{detunfide}

The state evolution under the secular approximation in Step-$1$ of Sec.~\ref{prepar} reflects the ideal assumption that the microwave driving pulses can only affect the desired pairs of states $\{n_1, m_1\}$ and $\{n_2, m_2\}$ filtered out by choosing the driving frequency in Eq.~(\ref{omega}). The two driving pulses might however induce the other state transitions that invalidate the secular-approximation condition $\Delta_{nm}, \Delta'_{nm}\gg\Omega$. Given the initial state $|\Phi(0)\rangle$ in Eq.~(\ref{initialstate}), the fidelity of the entangled state in Eq.~(\ref{phi}) is found to be
\begin{equation}\label{F}
F=\frac{\frac{\alpha^{2n_1}\beta^{2m_1}}{\sqrt{n_1!m_1!}}
+\frac{\alpha^{2n_2}\beta^{2m_2}}{\sqrt{n_2!m_2!}}}
{\sum_{n,m}\frac{\alpha^{2n}\beta^{2m}}{\sqrt{n!m!}}(P_{nm}+P'_{nm})}
\end{equation}
under both desired and undesired transitions. Here $E_{nm}=\sqrt{\Omega^2+(\Delta_{nm}/2)^2}$, $\tan(2\theta_{nm})=2\Omega/\Delta_{nm}$ and $P_{nm}=\sin^2(E_{nm}T)\sin^2(2\theta_{nm})$. $P'_{nm}$ is obtained by replacing $\Delta_{nm}$ with $\Delta'_{nm}$. Note $P_{n_1m_1}=P'_{n_2m_2}=1$. The detunings $\Delta_{nm}$ and $\Delta'_{nm}$ are given by Eq.~(\ref{delta12}). The derivation details can be found in Appendix~\ref{appb}.

This result is consistent with the condition that both $\Delta_{n\neq n_1,m\neq m_1}$ and $\Delta'_{n\neq n_2,m\neq m_2}$ should be sufficiently larger than $\Omega$. Then the undesired transitions are ignorable and does not obstruct generating the desired entangled state with a high fidelity. However, this condition cannot be always satisfied for arbitrary $|nm\rangle$. Similar to Eqs.~(\ref{omega}) and (\ref{delta12}), certain detunings $\Delta_{n'_1m'_1}$ and $\Delta'_{n'_2m'_2}$ approach vanishing when the Fock states indicated by $\{n'_1, m'_1\}$ and $\{n'_2, m'_2\}$ follow the same equations as the desired pairs of $\{n_1, m_1\}$ and $\{n_2, m_2\}$:
\begin{equation}\label{omega2}
\begin{aligned}
\omega_1&\approx\omega_f-\omega_e+(2n'_1+1)(\chi_a+\chi'_a)\\
&+m'_1(\chi_b+\chi'_b)+\chi_b,\\
\omega_2&\approx\omega_f+(2m'_2+1)(\chi_b+\chi'_b)+n'_2(\chi_a+\chi'_a)+\chi_a.
\end{aligned}
\end{equation}

If the coupling strength between qutrit and resonator is sufficiently weak, e.g., $g_{a,b}\sim0.01\omega_{a,b}$, then the energy shifts $\chi'_{a,b}$ induced by the counterrotating interactions $|f\rangle\langle e|a^{\dag}+$H.c. and $|f\rangle\langle g|b^{\dag}+$H.c. would be much less than the energy shifts $\chi_{a,b}$ induced by the rotating-wave interactions $|e\rangle\langle f|a^{\dag}+$H.c. and $|g\rangle\langle f|b^{\dag}+$H.c. as implied by Eq.~(\ref{chiab}). It is then interesting to find that Eq.~(\ref{omega2}) can be met under the conditions
\begin{equation}\label{fockratio}
\begin{aligned}
n'_1&=n_1+n_k,\quad n'_2=n_2+n'_k,\\
m'_1&=m_1-m_k,\quad m'_2=m_2-m'_k,\\
\frac{m_k}{2n_k}&\approx\frac{\chi_a}{\chi_b},\quad \frac{n'_k}{2m'_k}\approx\frac{\chi_b}{\chi_a},
\end{aligned}
\end{equation}
where $n_k$, $m_k$, $n'_k$, and $m'_k$ are nonnegative integers representing the shifts from the desired Fock states $\{n_1, m_1\}$ and $\{n_2, m_2\}$. In other words, the states $|en'_1m'_1\rangle$ ($|gn'_2m'_2\rangle$) would evolve to $|fn'_1m'_1\rangle$ ($|fn'_2m'_2\rangle$) in parallel to the desired transition $|en_1m_1\rangle\to|fn_1m_1\rangle$ ($|gn_2m_2\rangle\to|fn_2m_2\rangle$) with a nonnegligible probability under the same driving $H_d$ in Eq.~(\ref{Htotorigin}).

To reduce the probability of the undesired state transitions, one can properly choose $\chi_a$ and $\chi_b$ so that the populations on the states $|n'_1m'_1\rangle$ and $|n'_2m'_2\rangle$ become significantly less than those on $|n_1m_1\rangle$ and $|n_2m_2\rangle$. This requirement is satisfied when the ``state-shift'' ratios $m_k/n_k$ and $m'_k/n'_k$ are irreducible ratios of integers larger than the average number of excitations for the fixed initial coherent states, i.e., $|\alpha|^2$ or $|\beta|^2$. Due to Eq.~(\ref{fockratio}), the ``state-shift'' ratios are twice the energy-shift ratios. And according to Eq.~(\ref{chiab}), $\chi_a/\chi_b$ is determined by the transition frequencies of the qutrit and two resonators,
\begin{equation}\label{ratioab}
\frac{\chi_a}{\chi_b}=\frac{\omega_f-\omega_b}{\omega_f-\omega_e-\omega_a},
\end{equation}
when the coupling strengths are isotropic $g_a=g_b=g$. The ratio is thus an important element to the state fidelity, which can be properly adjusted by the frequencies of the system components. Its value should not be taken in the proximity of an irreducible ratio of small integers.

\begin{figure}[htbp]
\centering
\includegraphics[width=0.45\textwidth]{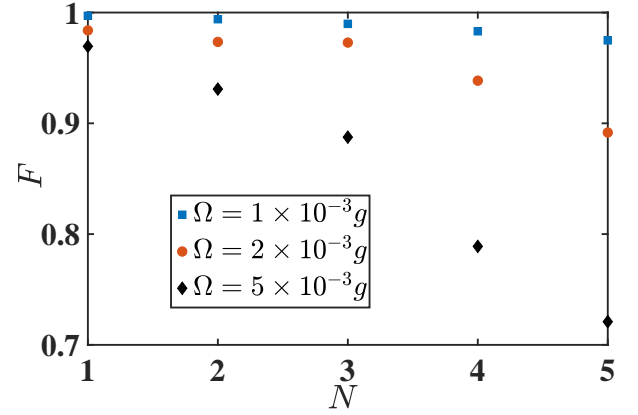}
\caption{Fidelity of the NOON state $(|0N\rangle+|N0\rangle)/\sqrt{2}$ as a function of $N$ under various Rabi frequencies $\Omega$. The parameters are set as $\alpha=\beta=1$, $g_a=g_b=g$, $\omega_e=20g$, $\omega_f=100g$, $\omega_a=70g$, and $\omega_b=89g$.}\label{noonfidelity}
\end{figure}

Figure~\ref{noonfidelity} confirms our two-step protocol in generating the NOON state from coherent states with $\alpha=\beta=1$. The state fidelity obtained by Eq.~(\ref{fidelity}) or Eq.~(\ref{F}) is plotted under various Rabi frequencies $\Omega$ of the driving pulses in the total Hamiltonian~(\ref{Htot}). For arbitrary $N$, $\Delta_{nm}, \Delta'_{nm}\gg\Omega$ would become invalid with a higher probability under a larger $\Omega$. For example, the fidelity of the NOON state with $N=2$ is $0.994$ when $\Omega=1\times10^{-3}g$ and it declines to $0.938$ when the Rabi frequency is enhanced to $\Omega=5\times10^{-3}g$. On the other hand, one can find that a larger $N$ gives rise to a smaller fidelity and the decline magnitude increases by increasing $\Omega$. In comparison to the state of $N=2$, the state fidelity for $N=5$ declines from $0.970$ ($\Omega=1\times10^{-3}g$) to $0.721$ ($\Omega=5\times10^{-3}g$).

\begin{figure}[htbp]
\centering
\includegraphics[width=0.45\textwidth]{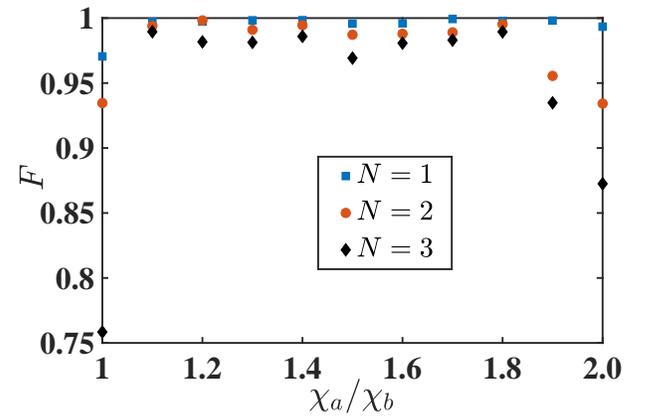}
\caption{Fidelity of the NOON state $(|0N\rangle+|N0\rangle)/\sqrt{2}$ as a function of the ratio of energy shifts $\chi_a/\chi_b$ under various $N$. The transition frequency $\omega_e$ is changed to manipulate $\chi_a/\chi_b$. The other
parameters are the same as Fig.~\ref{noonfidelity}. }\label{ratiofidelity}
\end{figure}

To avoid the participation of the undesired state transition involving $|n'_1m'_1\rangle$ and $|n'_2m'_2\rangle$, the ratio of $\chi_a/\chi_b$ in Fig.~\ref{noonfidelity} is set to be $11/10$. Both numerator and denominator are sufficiently distinct from either $|\alpha|^2$ or $|\beta|^2$. In Fig.~\ref{ratiofidelity}, we plot the NOON state fidelity as a function of $\chi_a/\chi_b$ in the range of $[1, 2]$. As expected, the fidelity declines when increasing the Fock number $N$. More importantly, one can observe that the fidelity declines significantly at certain values that could be expressed by irreducible ratio of small integers, e.g., $\chi_a/\chi_b=1$, $\chi_a/\chi_b=1.5$, and $\chi_a/\chi_b=2$. It can be theoretically explained by Eq.~(\ref{fockratio}). For example, $\chi_a/\chi_b=1$ gives rise to $m_k/n_k=2$. Under this condition, in parallel to the desired state transition $|e10\rangle\to|f10\rangle$, we have an undesired evolution described by $|e02\rangle\to|f02\rangle$ under the total Hamiltonian~(\ref{Htot}). Similarly, the populations on the states $|e12\rangle$ and $|e04\rangle$ cannot be ignored if the target transition is $|e20\rangle\to|f20\rangle$. In contrast, the fidelities are nearly unit at the other points, e.g., $\chi_a/\chi_b=1.1$ and $\chi_a/\chi_b=1.7$, where the population on $|en'_1m'_1\rangle$ satisfying the condition in Eq.~(\ref{fockratio}) is significant less than that on the target state $|en_1m_1\rangle$, regarding $|\alpha=1\rangle$. Thus, a high-fidelity entangled state can be achieved by properly choosing the initial coherent states of the resonators and the level spacings in ancillary qutrit.

\subsection{Systematic errors on driving pulses}\label{sysfide}

From Eqs.~(\ref{phit}) and (\ref{phiTt}), it is important for our protocol to perform the measurement at the keypoint $\Omega T=\pi/2$. Then it seems that the driving intensities $\Omega$ for both $|f\rangle\langle e|$ and $|f\rangle\langle g|$ have to take the same magnitude. And the driving frequencies $\omega_1$ and $\omega_2$ should be exactly determined by Eq.~(\ref{omega}) to realize the desired state evolution. The control might not be exactly implemented because of the technical imperfections and constrains. In this subsection, we consider the systematic errors raised by the unequal driving intensities and the deviations in driving frequencies.

We first consider that the two driving intensities depart from each other by a relative magnitude $\epsilon$. In particular, the driving Hamiltonian in Eq.~(\ref{Htotorigin}) is rewritten as
\begin{equation}
H_d=\Omega\left[(1-\epsilon)|f\rangle\langle e|e^{-i\omega_1t}+(1+\epsilon)|f\rangle\langle g|e^{-i\omega_2t}+{\rm H.c.}\right].
\end{equation}
and consequently the total Hamiltonian~(\ref{Htot}) in rotating frame becomes
\begin{equation}\label{Htotsys}
\begin{aligned}
&H_{\rm err}=\Omega(1-\epsilon)|fn_1m_1\rangle\langle en_1m_1|\\
+&\Omega(1+\epsilon)|fn_2m_2\rangle\langle gn_2m_2|\\
+&\Omega(1-\epsilon)\sum_{n\ne n_1,m\ne m_1}^{\infty}|fnm\rangle\langle enm|e^{i\Delta_{nm}t}\\
+&\Omega(1+\epsilon)\sum_{n\ne n_2,m\ne m_2}^{\infty}|fnm\rangle\langle gnm|e^{i\Delta'_{nm}t}+{\rm H.c.}.
\end{aligned}
\end{equation}
With the Hamiltonian in Eq.~(\ref{Htotsys}), the initial state $|\Phi(0)\rangle$ in Eq.~(\ref{initialstate}) evolves to
\begin{equation}\label{phiTtsys}
\begin{aligned}
&|\Phi(T)\rangle\\
&\approx\mathcal{N}\frac{\alpha^{n_1}\beta^{m_1}}{\sqrt{n_1!m_1!}}
\left[\sin\left(\frac{\pi\epsilon}{2}\right)|e\rangle-i\cos\left(\frac{\pi\epsilon}{2}\right)|f\rangle\right]|n_1m_1\rangle\\
&-\mathcal{N}\frac{\alpha^{n_2}\beta^{m_2}}{\sqrt{n_2!m_2!}}
\left[\sin\left(\frac{\pi\epsilon}{2}\right)|g\rangle+i\cos\left(\frac{\pi\epsilon}{2}\right)|f\rangle\right]|n_2m_2\rangle\\
&+\mathcal{N}\sum^{\infty}_{n\neq n_1,m\neq m_1}\frac{\alpha^n\beta^m}{\sqrt{n!m!}}|enm\rangle\\
&+\mathcal{N}\sum^{\infty}_{n\neq n_2,m\neq m_2}\frac{\alpha^n\beta^m}{\sqrt{n!m!}}|gnm\rangle
\end{aligned}
\end{equation}
when $\Omega T=\pi/2$. Then we measure the qutrit with the projection operator $M_f$, one can still obtain the (unnormalized) target state
\begin{equation}\label{phisys}
|\phi\rangle=\cos\left(\frac{\pi\epsilon}{2}\right)\left[\cos\varphi|n_1m_1\rangle+\sin\varphi|n_2m_2\rangle\right].
\end{equation}
Since $\cos(\pi\epsilon/2)$ is a common factor, the error $\epsilon$ on driving intensity can therefore be ignored in the ideal situation that the microwave pulses only drive the desired states. Its effect, however, emerges when considering the undesired evolution induced by vanishing $\Delta_{nm}$ and $\Delta'_{nm}$ as discussed in Sec.~\ref{detunfide}.

\begin{figure}[htbp]
\centering
\includegraphics[width=0.45\textwidth]{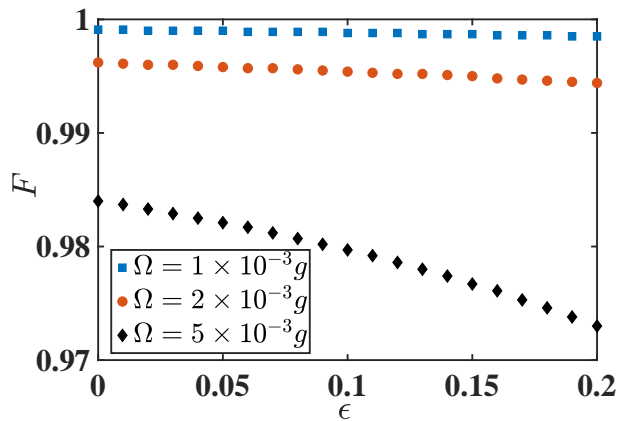}
\caption{Fidelity of the double-excitation Bell state $(|00\rangle+|11\rangle)/\sqrt{2}$ as a function of the intensity error $\epsilon$. The parameters are the same as Fig.~\ref{noonfidelity}.}\label{bellsysfidelity}
\end{figure}

\begin{figure}[htbp]
\centering
\includegraphics[width=0.45\textwidth]{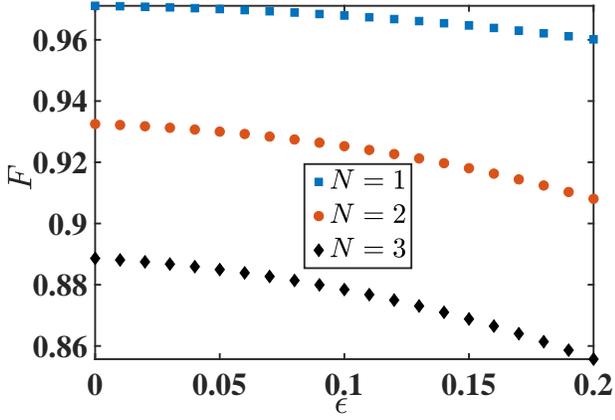}
\caption{Fidelity of the NOON state $(|0N\rangle+|N0\rangle)/\sqrt{2}$ as a function of the systematic error $\epsilon$ associated with the driving intensity under various $N$. $\Omega=5\times 10^{-3}g$ and the other parameters are the same as Fig.~\ref{noonfidelity}.}\label{sysfidelity}
\end{figure}

Under the Hamiltonian~(\ref{Htotsys}) with the intensity error, the fidelities of the double-excitation Bell state $(|00\rangle+|11\rangle)/\sqrt{2}$ and the NOON state $(|0N\rangle+|N0\rangle)/\sqrt{2}$ are presented in Fig.~\ref{bellsysfidelity} and Fig.~\ref{sysfidelity}, respectively, as a function of $\epsilon$. The fidelities are evaluated at the desired moment $\Omega T=\pi/2$ as requested by Step-$2$ in our protocol.

It is found in Fig.~\ref{bellsysfidelity} that the fidelity sensitivity to the intensity error $\epsilon$ is amplified with increasing Rabi frequency $\Omega$. In particular, for $\Omega=1\times 10^{-3}g$, the fidelities are roughly invariant in the presence of $\epsilon$, even when it is about $20\%$. And for $\Omega=5\times 10^{-3}g$, it declines slightly with $\epsilon$ when it increases from $0$ to $20\%$. This result could be understood since the condition $\Omega(1+\epsilon)\ll\chi_a,\chi_b$ becomes less reliable for a larger $\Omega$. In Fig.~\ref{sysfidelity}, the results for various $N$ are obtained under $\Omega=5\times 10^{-3}g$. Although a larger $N$ gives rise to more decay, the fidelities can still be maintained over $0.86$ with about $20\%$ errors in the driving intensity. Thus, one can conclude that our protocol is robust against the fluctuations on the Rabi frequencies. We can have a high fidelity as long as the measurement moment is determined by the average of these Rabi frequencies.

As for the deviation in driving frequency, we have
\begin{equation}\label{Hamderr}
H_d=\Omega\left[|f\rangle\langle e|e^{-i\omega_1(1+\epsilon')t}+|f\rangle\langle g|e^{-i\omega_2(1+\epsilon')t}+{\rm H.c.}\right].
\end{equation}
Here the relative errors for the two driving frequencies are assumed to take the same magnitude $\epsilon'$ with $\omega_j\epsilon'\ll\Omega$ and then the total Hamiltonian in the rotating frame with respect to $U(t)=\exp(iH_{\rm eff}t)$ is rewritten as
\begin{equation}\label{Htoterr}
\begin{aligned}
H_{\rm err}&=\Omega|fn_1m_1\rangle\langle en_1m_1| e^{-i\omega_1\epsilon' t}\\
&+\Omega|fn_2m_2\rangle\langle gn_2m_2|e^{-i\omega_2\epsilon' t}\\
&+\Omega\sum^{\infty}_{n\neq n_1,m\neq m_1}|fnm\rangle\langle enm|e^{i\Delta_{nm}t-i\omega_1\epsilon' t}\\
&+\Omega\sum^{\infty}_{n\neq n_2,m\neq m_2}|fnm\rangle\langle gnm|e^{i\Delta_{nm}t-i\omega_2\epsilon' t}+\rm{H.c.}.
\end{aligned}
\end{equation}
Consequently, the initial state $|\Phi(0)\rangle$ evolves to
\begin{equation}\label{phitott}
\begin{aligned}
|\Phi(T)\rangle&\approx\mathcal{N}\frac{\alpha^{n_1}\beta^{m_1}}{\sqrt{n_1!m_1!}}|\phi'_1\rangle
+\mathcal{N}\frac{\alpha^{n_2}\beta^{m_2}}{\sqrt{n_2!m_2!}}|\phi'_2\rangle\\
&+\mathcal{N}\sum^{\infty}_{n\neq n_1,m\neq m_1}\frac{\alpha^n\beta^m}{\sqrt{n!m!}}|enm\rangle\\
&+\mathcal{N}\sum^{\infty}_{n\neq n_2,m\neq m_2}\frac{\alpha^n\beta^m}{\sqrt{n!m!}}|gnm\rangle
\end{aligned}
\end{equation}
at the desired time $T=\pi/(2\Omega)$, where
\begin{equation}\label{phi12sys}
\begin{aligned}
|\phi'_1\rangle&=(\cos\vartheta_1+i\sin\vartheta_1\cos\theta_1)|en_1m_1\rangle\\
&-i\sin\vartheta_1\sin\theta_1|fn_1m_1\rangle,\\
|\phi'_2\rangle&=(\cos\vartheta_2+i\sin\vartheta_2\cos\theta_2)|gn_2m_2\rangle\\
&-i\sin\vartheta_2\sin\theta_2|fn_2m_2\rangle
\end{aligned}
\end{equation}
with $\vartheta_j\equiv\pi/(2\sin\theta_j)$ and $\tan\theta_j\equiv2\Omega/(\omega_j\epsilon')$. Then one can obtain
\begin{equation}\label{phitar}
|\phi'\rangle=\cos\varphi\sin\vartheta_1\sin\theta_1|n_1m_1\rangle+\sin\varphi\sin\vartheta_2\sin\theta_2|n_2m_2\rangle
\end{equation}
by performing the projective measurement $M_f$. Up to the second order of $\epsilon'$, it is found that
\begin{equation}\label{fidelityerr}
F=|\langle\phi|\phi'\rangle|^2\approx1-\frac{\omega^2_1\cos^2\varphi+\omega^2_2\sin^2\varphi}{4\Omega^2}\epsilon'^2.
\end{equation}

\begin{figure}[htbp]
\centering
\includegraphics[width=0.45\textwidth]{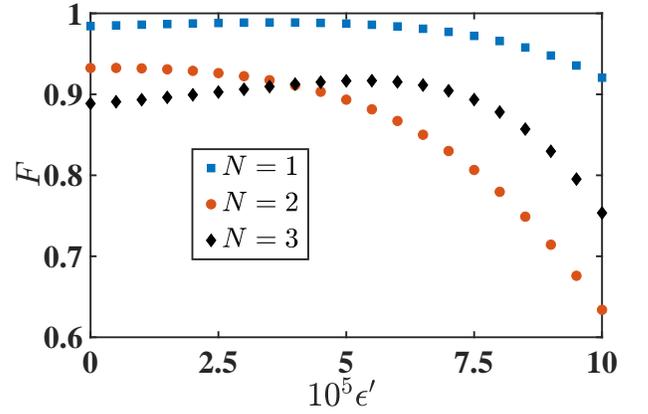}
\caption{Fidelity of the NOON state $(|0N\rangle+|N0\rangle)/\sqrt{2}$ as a function of the systematic error about the driving frequency $\epsilon'$ under various $N$. $\Omega=5\times 10^{-3}g$ and the other parameters are the same as  Fig.~\ref{noonfidelity}.}\label{syserrorfrequency}
\end{figure}

The fidelity sensitivity for the NOON state to the frequency error $\epsilon'$ is demonstrated in Fig.~\ref{syserrorfrequency}. For $N=1$, the fidelity declines slightly with $\epsilon'$ and it can be maintained over $0.92$ when the error magnitude approaches $\sim10^{-4}$. The decline rate becomes larger for $N=2$ and the fidelity is about $0.64$ when $\epsilon'=10^{-4}$. The behavior of fidelity for $N=3$ is not monotonic. It first increases until $\epsilon'\approx6\times10^{-5}$ and then gradually decreases to $0.75$ when $\epsilon'=10^{-4}$. In comparison to Fig.~\ref{sysfidelity}, our protocol is therefore less robust against the error in the driving frequency than that in the driving intensity.

\subsection{External decoherence}\label{decofide}

During the state generation, the composite system can not be completely isolated from the surrounding environment. The target entangled state will be damaged by the influence from both resonator damping and qutrit decoherence. Taking the local decoherence channels into consideration, we can study the entangled state generation fidelity in a standard open-quantum-system framework. Under the Markovian approximation and tracing out the degrees of freedom of the external environments (assumed to be at the vacuum states), we arrive at the master equation for the density operator $\rho(t)$ for the composite system
\begin{equation}\label{lindblad}
\begin{aligned}
\dot{\rho}(t)&=-i\left[H'_{\rm tot}, \rho(t)\right]
+\gamma_{fe}\mathcal{L}[|e\rangle\langle f|]+\gamma_{fg}\mathcal{L}[|g\rangle\langle f|]\\
&+\gamma_e\mathcal{L}[|e\rangle\langle e|]+\gamma_f\mathcal{L}[|f\rangle\langle f|]+\kappa_a\mathcal{L}[a]+\kappa_b\mathcal{L}[b].
\end{aligned}
\end{equation}
Here $H'_{\rm{tot}}$ is the total Hamiltonian in Eq.~(\ref{Htot}). $\gamma_{fe}$ and $\gamma_{fg}$ are the population relaxation constants associated with the transitions $|f\rangle\to|e\rangle$ and $|f\rangle\to|g\rangle$, respectively. $\gamma_e$ and $\gamma_f$ are the dephasing rates associated with the states $|e\rangle$ and $|f\rangle$, respectively. For simplicity, these decoherence rates of qutrit are assumed to take the same value $\gamma$. $\kappa_a$ ($\kappa_b$) is the decay constant of the resonator mode $a$ ($b$). We set $\kappa_a=\kappa_b=0.1\gamma$ for the resonator lifetime is higher than the qutrit lifetime in realistic experimental platforms~\cite{neardelta,resonator,circuitqed}. The Lindblad superoperator $\mathcal{L}$ is defined as
\begin{equation}\label{supop}
\mathcal{L}[o]\equiv o\rho o^\dag-\frac{1}{2}o^\dag o\rho-\frac{1}{2}\rho o^\dag o.
\end{equation}

\begin{figure}[htbp]
\centering
\includegraphics[width=0.45\textwidth]{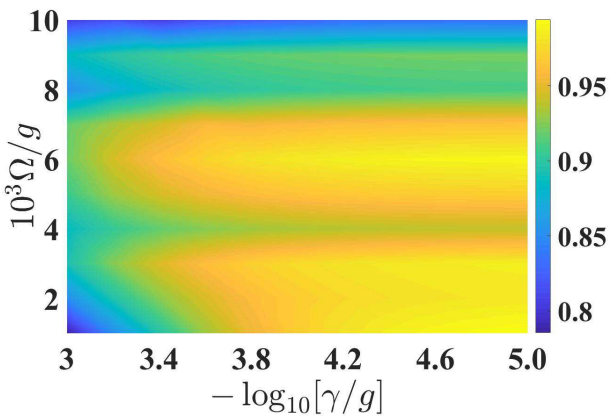}
\caption{Fidelity of the Bell state $(|01\rangle+|10\rangle)/\sqrt{2}$ in the parametric space of Rabi frequency $\Omega$ and the decoherence rate $\gamma$. The other parameters are the same as Fig.~\ref{noonfidelity}.}\label{masterfide}
\end{figure}

The fidelity of the target state $(|01\rangle+|10\rangle)/\sqrt{2}$ under various Rabi frequencies $\Omega$ and decoherence rates $\gamma$ is shown in Fig.~\ref{masterfide}. In the presence of a larger decoherence rate $\gamma$ (the left region in the figure), a smaller Rabi frequency $\Omega$ cannot yield a higher fidelity as in Figs.~(\ref{noonfidelity}) and (\ref{bellsysfidelity}). We have to make a compromise of $\Omega$ to optimize the fidelity since the desired period $T$ of the state generation is inversely proportional to the $\Omega$ and the decoherence becomes more severe under a longer evolution time. It is found that the fidelity in two optimized regimes is maintained above $0.95$. One is around $5\times10^{-3}g<\Omega<7\times10^{-3}g$ even when $\gamma$ is nearly $10^{-3.5}g$; another one is around $1\times10^{-3}g<\Omega<3\times10^{-3}g$ when $\gamma<10^{-3.8}g$.

\begin{figure}[htbp]
\centering
\includegraphics[width=0.45\textwidth]{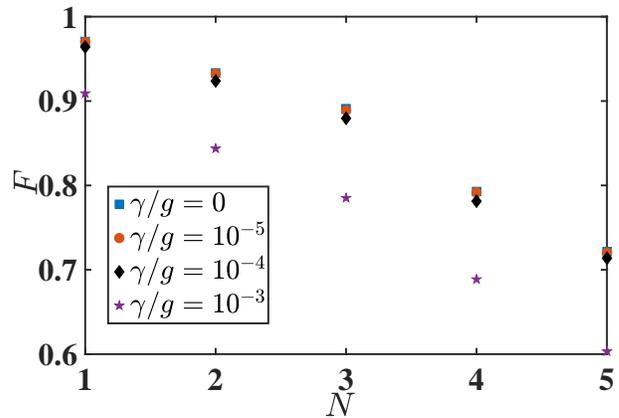}
\caption{Fidelity of the NOON state $(|0N\rangle+|N0\rangle)/\sqrt{2}$ as a function of $N$ under various decoherence rates. $\Omega=5\times 10^{-3}g$ and the other parameters are the same as Fig.~\ref{noonfidelity}.}\label{masterfidelity}
\end{figure}

In Fig.~\ref{masterfidelity}, we present the impact of decoherence on the NOON state fidelity under $\Omega=5\times 10^{-3}g$. The fidelity declines with the increasing Fock number $N$ and it is not sensitive to the decoherence rate when $\gamma/g\leq10^{-4}$. For $\gamma/g=10^{-4}$, the fidelities are $0.964$, $0.924$, $0.880$, $0.781$, and $0.714$ for $N=1,2,3,4,5$, respectively. A significant decline in fidelity occurs under a stronger decoherence rate $\gamma/g=10^{-3}$. It is about $0.603$ for $N=5$.

For such a system of a flux qutrit and two resonators~\cite{fq,fq3,sq,sq3}, the transition frequencies among the lowest three levels of the qutrit can be manipulated within the range of $[1,20]$ GHz, and the frequencies of the resonators are in the order of $\sim 10$ GHz. The coupling between qutrit and resonators has entered the strong and even ultrastrong coupling regimes~\cite{sq,sq2}, i.e., $g/\omega\sim0.01-0.1$. The coherence timescale of the qutrit is about $\gamma^{-1}\sim10 \mu s$~\cite{neardelta}, corresponding to $\gamma/\omega\sim10^{-5}$. The quality factor of the resonators is about $Q=\omega/\kappa\sim 10^6-10^7$~\cite{circuitqed}, and even approaches $10^8$~\cite{resonator}, corresponding to $\kappa/\omega\sim10^{-7}-10^{-6}$. Then we set $\gamma/g\sim10^{-5}-10^{-3}$ and $\kappa=0.1\gamma$ in preceding calculations.

\subsection{Unwanted couplings}\label{uncoupling}

In addition to the external noises, our protocol could be also subject to the unwanted couplings inside the system. In Sec.~\ref{model}, we use a $\Lambda$-type qutrit to approximate the realistic $\Delta$-type qutrit, depending on the suppressed dipole-dipole transition between $|e\rangle$ and $|g\rangle$. This transition is however unavoidable under $\Phi/\Phi_0\neq0.5$ [see Fig.~\ref{diagram1}(c)]. In this case, the interaction Hamiltonian $V$ in the initial Hamiltonian~(\ref{deltaHamiltonian}) is then rewritten as
\begin{equation}\label{noonmodel}
\begin{aligned}
\tilde{V}&=[g_a(a^{\dag}+a)+g_b(b+b^{\dag})]\times \\
&(|e\rangle\langle g|+|g\rangle\langle e|+|f\rangle\langle e|+|e\rangle\langle f|+|f\rangle\langle g|+|g\rangle\langle f|).
\end{aligned}
\end{equation}
To simplify the discussion but with no loss of generality, it is assumed that $g^n_{eg}=g^n_{fg}=g^n_{fe}=g_n$, $n=a,b$. With the second-order perturbation method presented in Eq.~(\ref{chi}), the effective Hamiltonian in Eq.~(\ref{Heff}) for the whole system is modified to
\begin{equation}\label{Hefftot}
\begin{aligned}
\tilde{H}_{\rm eff}&=\sum_{k,j\in{g,e,f}; l\in{a,b}}\left[\omega_k+\Theta(kj)\chi^l_{kj}\right]|k\rangle\langle k|\\
&+\sum_{k,j\in{g,e,f}; l\in{a,b}}\chi^l_{kj}l^\dag l\Theta(kj)\left(|k\rangle\langle k|-|j\rangle\langle j|\right),
\end{aligned}
\end{equation}
where the energy shifts read
\begin{equation}
\chi^l_{kj}=\frac{g^2_l}{\Theta(kj)\left(\omega_k-\omega_j-\omega_l\right)}.
\end{equation}
We here employ the step function $\Theta(kj)$, i.e., $\Theta(kj)=-1$ when $\omega_k<\omega_j$ and $\Theta(kj)=1$ when $\omega_k>\omega_j$.

The driving Hamiltonian $H_d$ in Eq.~(\ref{Htotorigin}) is still applicable to distinguish the desired state transitions $|en_1m_1\rangle\to|fn_1m_1\rangle$ and $|gn_2m_2\rangle\to|fn_2m_2\rangle$. Under the modified effective Hamiltonian in Eq.~(\ref{Hefftot}), it is found that the two frequencies of microwave pulses in Eq.~(\ref{omega}) are respectively modified to
\begin{equation}\label{totalfrequency}
\begin{aligned}
\tilde{\omega}_1&=\omega_f-\omega_e+(2n_1+1)(\chi^a_{ef}+\chi^a_{fe})+n_1(\chi^a_{gf}-\chi^a_{ge})\\
&+(n_1+1)(\chi^a_{fg}-\chi^a_{eg})+(2m_1+1)(\chi^b_{ef}+\chi^b_{fe})\\
&+m_1(\chi^b_{gf}-\chi^b_{ge})+(m_1+1)(\chi^b_{fg}-\chi^b_{eg}),\\
\tilde{\omega}_2&=\omega_f+(2n_2+1)(\chi^a_{gf}+\chi^a_{fg})+n_2(\chi^a_{ef}+\chi^a_{eg})\\
&+(n_2+1)(\chi^a_{fe}+\chi^a_{ge})+(2m_2+1)(\chi^b_{gf}+\chi^b_{fg})\\
&+m_2(\chi^b_{ef}+\chi^b_{eg})+(m_2+1)(\chi^b_{fe}+\chi^b_{ge}).
\end{aligned}
\end{equation}
The total Hamiltonian in analog to Eq.~(\ref{Htotorigin}) is then written as
\begin{equation}\label{Htotlab}
\tilde{H}_{\rm tot}=\tilde{H}_{\rm eff}+H_d.
\end{equation}
Subsequently one can carry out a similar rotation as in Step-$1$ in our protocol to obtain the desired time-independent Hamiltonian in Eq.~(\ref{Htot}). Under the secular approximation, it is used to push the system into the state in Eq.~(\ref{phiTt}) for Step-$2$.

It is straightforward to check that the transition frequencies $\omega_1$ and $\omega_2$ in Eq.~(\ref{omega}) could be respectively recovered by $\tilde{\omega}_1$ and $\tilde{\omega}_2$ when the energy shifts $\chi^a_{fe}\gg\chi^a_{eg}, \chi^a_{fg}$ and $\chi^b_{fg}\gg\chi^b_{eg}, \chi^b_{fe}$. It is consistent with the fact that the $\Delta$-type qutrit could be approximated by the $\Lambda$-type qutrit in Sec.~\ref{model}. In this case, $\chi^a_{fe}=\chi_a$, $\chi^a_{ef}=\chi'_a$, $\chi^b_{fg}=\chi_b$, and $\chi^b_{gf}=\chi'_b$, where $\chi_a$, $\chi'_a$, $\chi_b$, and $\chi'_b$ have been given in Eq.~(\ref{chiab}).

\begin{figure}[htbp]
\centering
\includegraphics[width=0.45\textwidth]{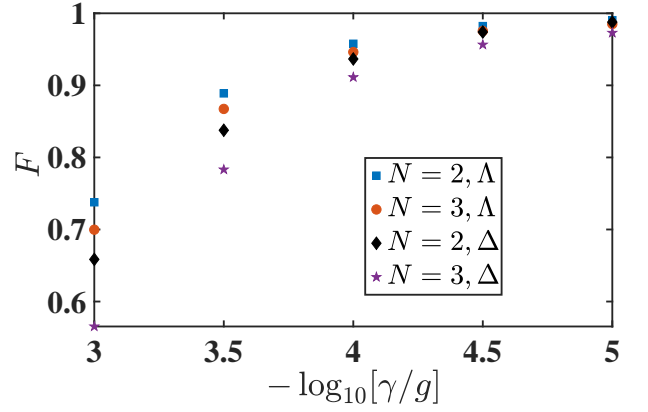}
\caption{Fidelity of the NOON state $(|0N\rangle+|N0\rangle)/\sqrt{2}$ with $N=2,3$ as a function of the decoherence rate $\gamma$ for the $\Lambda$-type qutrit under the total Hamiltonian $H_{\rm tot}$ in Eq.~(\ref{Htotorigin}) and the $\Delta$-type one under $\tilde{H}_{\rm tot}$ in Eq.~(\ref{Htotlab}). $\Omega=1\times 10^{-3}g$ and the other parameters are the same as Fig.~\ref{masterfidelity}.}\label{experfide}
\end{figure}

In Fig.~\ref{experfide}, we present the NOON state fidelity by using the Lindblad master equation~(\ref{lindblad}) with $H_{\rm tot}$ for the $\Lambda$-type qutrit and $\tilde{H}_{\rm tot}$ for the $\Delta$-type one. The extra decoherence channel described by $\mathcal{L}[|g\rangle\langle e|]$ has been taken account into Eq.~(\ref{lindblad}) with the same dissipation rate $\gamma$ when we consider the $\Delta$-type qutrit. We find that the discrepancy between the $\Lambda$-type and the $\Delta$-type is enlarged by the decoherence rate. For example, for $N=2$, the fidelity of the $\Lambda$-type qutrit is $0.990$ under $\gamma/g=10^{-5}$, which is almost the same as that of the $\Delta$-type $0.988$. In contrast, the fidelity under $\gamma/g=10^{-3}$ declines to $0.738$ and $0.659$ for the $\Lambda$-type and the $\Delta$-type, respectively. Note the decoherence rates in practice satisfy $\gamma_{fg}, \gamma_{fe}\gg\gamma_{eg}$~\cite{sq3}. Our calculation about the $\Delta$-type qutrit could thus be regarded as a lowerbound estimation.

\begin{figure}[htbp]
\centering
\includegraphics[width=0.45\textwidth]{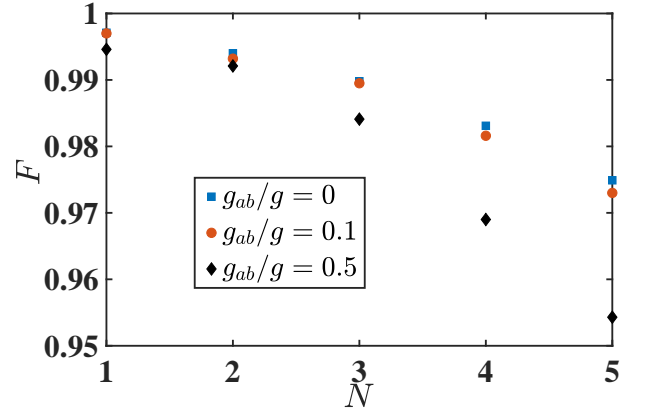}
\caption{Fidelity of the NOON state $(|0N\rangle+|N0\rangle)/\sqrt{2}$ as a function of $N$ under various strength of crosstalk $g_{ab}$ between two resonators. $\Omega=1\times 10^{-3}g$ and the other parameters are the same as Fig.~\ref{noonfidelity}.}\label{gabfidelity}
\end{figure}

Another unwanted coupling arises from the crosstalk between the two resonators $a$ and $b$. In this situation, the interaction Hamiltonian $V$ in Eq.~(\ref{Hamiltonian}) becomes
\begin{equation}\label{Vsgab}
\tilde{V}=V+g_{ab}\left(a+a^{\dag}\right)\left(b+b^{\dag}\right),
\end{equation}
where $g_{ab}$ is the inter-resonator coupling strength. Then the total Hamiltonian in Eq.~(\ref{Htotorigin}) turns out to be
\begin{equation}\label{Htotcross}
\tilde{H}_{\rm tot}=H_{\rm tot}+g_{ab}(a+a^{\dag})(b+b^{\dag}).
\end{equation}

In Fig.~\ref{gabfidelity}, the NOON state fidelities under various crosstalk coupling strength $g_{ab}$ are calculated in the absence of decoherence. The crosstalk is also found to be a negative ingredient to the fidelity of the generated state for any $N$. Yet its effect is not significant. For $N=5$, comparing to $F=0.975$ when $g_{ab}=0$, the fidelity can be maintained as $F=0.954$ even when $g_{ab}=0.5g$.

\section{Discussion}\label{discuss}

In addition to the $\Delta$-type qutrit for $\Phi/\Phi_0\neq0.5$ as discussed in Sec~\ref{uncoupling}, our protocol can be implemented by a $\Xi$-type transmon qutrit with a specific ratio $\Phi/\Phi_0=0.5$ [see Fig.~\ref{diagram1}(b)]. In this case, the total Hamiltonian in Eq.~(\ref{deltaHamiltonian}) could be rewritten as
\begin{equation}\label{Hamxi}
\begin{aligned}
H&=H_0+V,\\
H_0&=\omega_a a^{\dag}a+\omega_b b^{\dag}b+\omega_e|e\rangle\langle e|+\omega_f|f\rangle\langle f|,\\
V&=g_a(|e\rangle\langle g|+|g\rangle\langle e|)(a+a^{\dag})\\
&+g_b(|f\rangle\langle e|+|e\rangle\langle f|)(b+b^\dag).
\end{aligned}
\end{equation}
In the absence of the crosstalk between the two resonators, the effective Hamiltonian under the second-order perturbation~(\ref{chi}) becomes
\begin{equation}\label{Heffci}
\begin{aligned}
H_{\rm{eff}}&=-\chi_a'|g\rangle\langle g|+(\omega_e+\chi_a-\chi'_b)|e\rangle\langle e|\\
&+(\omega_f+\chi_b)|f\rangle\langle f|+\omega_a a^{\dag}a+\omega_b b^\dag b\\
&+(\chi_a+\chi'_a)a^{\dag}a(|e\rangle\langle e|-|g\rangle\langle g|)\\
&+(\chi_b+\chi'_b)b^\dag b(|f\rangle\langle f|-|e\rangle\langle e|),
\end{aligned}
\end{equation}
where
\begin{equation}\label{chiabxi}
\begin{aligned}
\chi_a&=\frac{g^2_a}{\omega_e-\omega_a}, \quad \chi'_a=\frac{g^2_a}{\omega_e+\omega_a},\\
\chi_b&=\frac{g^2_b}{\omega_f-\omega_e-\omega_b},  \quad  \chi'_b=\frac{g^2_b}{\omega_f-\omega_e+\omega_b}.
\end{aligned}
\end{equation}
Similar to Eq.~(\ref{Heff}), this effective Hamiltonian again describes the photon-number-dependent Stark shifts for each energy levels of qutrit. Following the same procedure illustrated in Sec.~\ref{prepar}, we can create an entangled state from the initial state $(|g\rangle+|f\rangle)/\sqrt{2}\otimes|\alpha\rangle|\beta\rangle$ by using tailored pulses transferring the states $|gn_1m_1\rangle\to|en_1m_1\rangle$ and $|fn_2m_2\rangle\to|en_2m_2\rangle$ and then measuring the qutrit with the projection operator $M_e\equiv|e\rangle\langle e|$. Also we can apply the Lindblad master equation (\ref{lindblad}) to estimate the entangled-state fidelities using a $\Xi$-type qutrit under decoherence, where the dissipation channel $\mathcal{L}[|g\rangle\langle f|]$ is now replaced by $\mathcal{L}[|g\rangle\langle e|]$.

\begin{table}[htbp]
\begin{tabular}{|c|c|c|c|c|c|}
  \hline
  $N$ & 1 & 2 & 3 & 4 & 5 \\
  \hline
  $F_{\Lambda}$ & 0.964 & 0.924& 0.880 & 0.781 & 0.714 \\
  \hline
  $F_{\Xi}$ & 0.917 & 0.891 & 0.736 & 0.412 & 0.124 \\
  \hline
\end{tabular}
\caption{Fidelities $F_{\Lambda}$ and $F_{\Xi}$ of the NOON state $(|0N\rangle+|N0\rangle)/\sqrt{2}$ using a $\Lambda$-type qutrit under $\gamma=1\times10^{-4}g$ and a $\Xi$-type one under $\gamma=5\times 10^{-5}g$, respectively~\cite{neardelta,entangle2}. Both driving intensities are set as $\Omega=5\times 10^{-3}g$. The other parameters are the same as Fig.~\ref{noonfidelity}, except in the $\Xi$-type qutrit $\omega_e=80g$ and $\omega_f=180g$ to comply the near-resonant condition with the two resonators.}\label{ftable}
\end{table}

In Table.~\ref{ftable}, we find that all fidelities with the $\Xi$-type qutrit are smaller than those with the $\Lambda$-type one, although the decoherence rate of the former is half of the latter. And their difference is enlarged by $N$. In our main protocol with the $\Lambda$-type qutrit, where the initial state is a superposed state of the two lowest levels $|g\rangle$ and $|e\rangle$. No extra decoherence channel populates the highest level $|f\rangle$ that serves as the measured state. While in the protocol with the $\Xi$-type qutrit, the initial state is a superposed state of $|g\rangle$ and $|f\rangle$. The measured middle level $|e\rangle$ is under a severe influence from the population relaxation by $\mathcal{L}[|e\rangle\langle f|]$. The final state fidelity with the $\Xi$-type qutrit is therefore not as high as that with the $\Lambda$-type one.

\section{Conclusion}\label{conclu}

In summary, we have presented a concise two-step protocol for creating an arbitrary bipartite entangled state in Fock space of two microwave resonators that are strongly coupled to a superconducting qutrit. No extra steps have to be taken to shape the initial states of the resonators. With a second-order perturbation method, the system could be described by an effective Hamiltonian in the dispersive regime. We take advantage of the excitation-number-dependent Stark shift for the qutrit transition frequency. It allows us to apply tailored microwave drive signals to individually control the qutrit transition amplitudes associated with the desired Fock states. Then an entangled states of the two resonators could be generated by a typical evolution-and-measurement procedure, merely from their initial coherent states.

The undesired state transitions induced by the vanishing detunings are the main obstacles to achieve a high-fidelity entangled state in our protocol. They could be properly suppressed by choosing the level spacings of system components in the presence of a strong driving. Moreover, our protocol is found to be robust against the systematic errors arising from the microwave driving intensity, the quantum decoherence of all components, and the crosstalk of two resonators. Our protocol can be conveniently extended to the qutrits in both $\Delta$ and $\Xi$ configurations. Hence, our study is of interest in the pursuit of Bell states and NOON states of continuous-variable systems with the dispersive interaction and finds important applications in state manipulation over the circuit-QED system with few steps.

\section*{Acknowledgments}

We acknowledge financial support from the National Science Foundation of China (Grant No. 11974311).

\appendix

\section{Nonideal measurement}\label{appa}

In our protocol with the $\Lambda$-type qutrit, the projective measurement $M_f$ is imposed on $|f\rangle$ to yield the desired entangled state. For a nonideal measurement, however, the projection operator becomes $M_{f'}=|f'\rangle\langle f'|$, where $|f'\rangle=\cos\tilde{\theta}_1|f\rangle+\sin\tilde{\theta}_1\sin\tilde{\theta}_2|e\rangle
+\sin\tilde{\theta}_1\cos\tilde{\theta}_2|g\rangle$ with $\tilde{\theta}_1\ll1$ and $\tilde{\theta}_2\in[0, \pi]$. The perfect measure is recovered when $\tilde{\theta}_1=0$. Using Eq.~(\ref{fidelity}), it is straightforward to find that for a balanced superposed entangled state ($\tan\varphi=1$), the fidelity becomes
\begin{equation}\label{fidemeasure}
\begin{aligned}
&F'=\frac{\langle\phi|\langle f'|\rho(T)|f'\rangle|\phi\rangle}{{\rm Tr}[M_{f'}\rho(T)M_{f'}]}\\
&\approx\frac{P(1+\tilde{\theta}^2_1/4)}{P(1+\tilde{\theta}^2_1/2)
+\mathcal{N}^2\sum^{\infty}_{n\neq n_1,n_2;m\neq m_1,m_2}\frac{\alpha^{2n}\beta^{2m}}{n!m!}\tilde{\theta}^2_1}\\
&\approx1-\left(\frac{1}{P}-\frac{3}{4}\right)\tilde{\theta}^2_1.
\end{aligned}
\end{equation}
Here $\mathcal{N}$ and $P$ are given by Eqs.~(\ref{initialstate}) and (\ref{succproba}), respectively. The deviation caused by nonideal measurement is therefore proportional to $\tilde{\theta}^2_1$ in the leading order.

\section{Fidelity under undesired state transitions}\label{appb}

In this appendix, we calculate the state fidelity~(\ref{F}) by considering the undesired evolution induced by vanishing $\Delta_{nm}$ and $\Delta'_{nm}$. The total Hamiltonian~(\ref{Htot}) is block-diagonal in terms of the subspaces spanned by $\{|fnm\rangle, |enm\rangle\}$ and $\{|fnm\rangle, |gnm\rangle\}$ with arbitrary pairs of Fock state ${|nm\rangle}$. And then in any subspace of $\{|fnm\rangle, |enm\rangle\}$, it can be effectively described (through rotating to a frame with time-independent coefficients) by
\begin{equation}\label{Htotmatrix}
H_{\rm loc}=\begin{bmatrix}
\Delta_{nm} & \Omega\\
\Omega & 0
\end{bmatrix}=\frac{\Delta_{nm}}{2}\hat{I}+\begin{bmatrix}
\frac{\Delta_{nm}}{2} & \Omega\\
\Omega & -\frac{\Delta_{nm}}{2}
\end{bmatrix},
\end{equation}
where $\hat{I}$ is a two-dimensional identity operator and $\Delta_{nm}$ is already given by Eq.~(\ref{delta12}). The local time evolution operator reads,
\begin{equation}\label{Ut}
U=\cos(E_{nm}t)\hat{I}-i\sin(E_{nm}t)\begin{bmatrix}
\cos(2\theta_{nm}) & \sin(2\theta_{nm})\\
\sin(2\theta_{nm}) & -\cos(2\theta_{nm})
\end{bmatrix},
\end{equation}
where $E_{nm}=\sqrt{\Omega^2+(\Delta_{nm}/2)^2}$, and $\tan(2\theta_{nm})=2\Omega/\Delta_{nm}$. Then for the initial state $|\psi(0)\rangle=|enm\rangle$, we have
\begin{equation}\label{varphit}
\begin{aligned}
|\psi(t)\rangle&=-i\sin(E_{nm}t)\sin(2\theta_{nm})|fnm\rangle\\
+&\left[\cos(E_{nm}t)+i\sin(E_{nm}t)\cos(2\theta_{nm})\right]|enm\rangle.
\end{aligned}
\end{equation}
At the measurement time $\Omega T=\pi/2$ in Step-$2$ of our protocol, the population on the number state $|nm\rangle$ of the resonators is found to be $P_{nm}=\sin^2(E_{nm}T)\sin^2(2\theta_{nm})$.

The time evolution in the subspace $\{|fnm\rangle, |gnm\rangle\}$ can be calculated in a similar way. Then the population on the state $|nm\rangle$ at the time $t=T=\pi/(2\Omega)$ is $P'_{nm}=\sin^2(E'_{nm}T)\sin^2(2\theta'_{nm})$, where $E'_{nm}$ and $\tan(2\theta'_{nm})$ are obtained through Eq.~(\ref{Ut}) by replacing $\Delta_{nm}$ with $\Delta'_{nm}$.

Given the initial state $|\Phi(0)\rangle$ in Eq.~(\ref{initialstate}), which is across all those separable subspaces, the fidelity of the desired entangled-state indicated by $\{n_1, m_1\}$ and $\{n_2, m_2\}$ is found to be
\begin{equation}\label{Fappb}
F=\frac{\frac{\alpha^{2n_1}\beta^{2m_1}}{\sqrt{n_1!m_1!}}
+\frac{\alpha^{2n_2}\beta^{2m_2}}{\sqrt{n_2!m_2!}}}
{\sum_{n,m}\frac{\alpha^{2n}\beta^{2m}}{\sqrt{n!m!}}(P_{nm}+P'_{nm})}
\end{equation}
according to Eq.~(\ref{fidelity}).

\bibliographystyle{apsrevlong}
\bibliography{reference}

\end{document}